\documentclass[preprint2]{emulateapj}

\usepackage{amssymb}
\usepackage{multirow}

\usepackage{graphicx}
\usepackage{enumerate}
\usepackage{amsmath}
\usepackage{enumitem}

\def\cm{\,{\rm cm}}

\def\ergscm2 {erg\,s$^{-1}$cm$^{-2}$}

\def\cm2 {cm$^{-2}$}

\def\aap {A\&A}
\def\apj {ApJ}

\shorttitle{SN 2014J:  Hints of a compact remnant}
\shortauthors{Ouyed et al.}

\begin{document}
 
\title{The puzzling early detection of low velocity $^{56}$Ni decay lines in SN 2014J:  Hints of a compact remnant}

\author{Rachid Ouyed\thanks{Email:rouyed@ucalgary.ca}, Denis Leahy, Nico Koning}
\affil{Department of Physics and Astronomy, University of Calgary, 2500 University Drive NW, Calgary, Alberta, T2N 1N4 Canada}
\and
\author{Jan Staff}
\affil{Department of Physics and Astronomy, Macquarie University, Sydney, Australia}

\begin{abstract}
We show that the  low-velocity $^{56}$Ni decay  lines detected earlier than 
expected in the type Ia SN 2014J find an explanation in the Quark-Nova Ia model which
 involves the thermonuclear explosion of  a tidally disrupted sub-Chandrasekhar White Dwarf in a tight 
 Neutron-Star-White-Dwarf  binary system.  
  The explosion is triggered by impact from the  Quark-Nova ejecta on the   WD material;
 the Quark-Nova is the explosive transition  of the Neutron star to a Quark star triggered  by accretion from  a
  CO torus (the circularized  WD material). 
   The presence of a compact remnant (the Quark Star) provides: (i) an additional energy source
   (spin-down power) which allows us to fit the observed light-curve including the 
   steep  early rise;   (ii) a central gravitational potential which 
    slows down some of the $^{56}$Ni produced   to velocities of a few $10^3$ km s$^{-1}$.
    In our model,  the $^{56}$Ni decay lines  become optically visible  
     at $\sim 20$ days from explosion time in agreement with observations.   
     We list predictions that can provide important
   tests for our model.
\end{abstract}


\section{Introduction}

SN 2014J  was discovered on Jan 21 2014 (Fossey et al. 2014)
 in M82 at 3.5 Mega-parsecs, making it one of the closest SNe-Ia observed in recent decades.
 Follow-up observations suggest that SN 2014J is a normal SN Ia (Goobar et al. 2014; 
 Ayani 2014; Cao et al. 2014; Itoh et al. 2014) and seems  consistent with a delayed-detonation explosion model (Marion et al.
 2014).   Pre-explosion optical images  of SN 2014J find no evidence for red supergiant companion stars (Goobar et al. 2014; Kelly et al. 2014), while non-detections in pre-explosion X-ray images seem inconsistent with the progenitor system being in a super-soft state  just before explosion (Maksym et al. 2014; Nielsen et al. 2014).
 Arguments for and against the single-degenerate (SD; Whelan \& Iben 1973) and the double-degenerate (DD; Iben \& Tutukov 1984) scenarios  have been  presented in the literature (e.g. Nielsen et al. 2014; see also Diehl et al. 2014a and references therein). 
  Churazov et al. (2014) observed SN 2014J at 50-100 days past explosion via $^{56}$Co
decay lines and  derived a visible $^{56}$Co mass
of $\sim 0.2M_{\odot}$ which translates to $\sim 0.36M_{\odot}$ initial $^{56}$Ni mass. 
They find a measured   $^{56}$Co expansion velocity of  a few $10^3$ km s$^{-1}$ and argued for
 a standard explosion by comparing 
 the  observed line shape  to their  ``fiducial" Chandrasekhar-mass model (their Figure 4).

A very surprising aspect of SN 2014J is the detection of  158 keV and 812 keV $^{56}$Ni
decay lines only $\sim 20$ days after the explosion  (Diehl et al. 2014a).  The corresponding $< 2000$
km s$^{-1}$ velocities are much lower than those measured in the optical ($\sim 10^4$ km s$^{-1}$; Marion et al. 2014). 
These detections were so surprising and puzzling  that Diehl et al. (2014) had
to consider a model  involving a Helium(He) belt with 
  the aim of confining  the resulting $^{56}$Ni ashes to 
 the equatorial plane.  This  picture might   account for the low velocities of the
 $^{56}$Ni decay lines if the  belt  is observed pole-on
 (their Figure 1).  Besides the strong constraint on the viewing angle, 
 there is no apparent mechanism 
  to constrain it from   spreading  isotropically rather than being confined to the equatorial plane.

    Here we present an alternative scenario involving a tight Neutron-Star-(CO)White-Dwarf (NS-COWD)\footnote{The system formed through a Common Envelope (CE) phase between the NS and the CO WD progenitor. During this phase the envelope is ejected leaving the NS-COWD system free to evolve to a tighter
     orbit  by gravitational   waves emission (Ouyed et al. 2014a).}  binary where the NS experiences an explosive transition to a quark star (QS):  the Quark-Nova (QN; Ouyed et al. 2002; Ker\"anen et al. 2005; Niebergal et al. 2010). The Quark-Nova Ia (QN-Ia) is the  thermonuclear
    explosion of the WD  material following impact by the relativistic and very dense QN ejecta.    A QN-Ia light curve is powered
     by two sources of energy: the $^{56}$Ni decay energy and the spin-down energy from the QS.
    The  QN-Ia has 
 been studied in previous papers (Ouyed\&Staff 2013; Ouyed et al. 2014a; Ouyed et al. 2014b; see  Ouyed et al. 2011
  for QNe in Low-Mass X-ray Binaries in general) where the
 interested reader can find details.   As we argue in this paper, the QN-Ia model  provides a reasonable account of the observed features of the $^{56}$Ni decay  lines
    in SN 2014J namely:
(i) the low expansion velocities; (ii) the low optical depth 
 at $\sim 20$ days from explosion; (iii) the amount of $^{56}$Ni produced; (iv)  the  light-curve, including the early steep
rise.

\begin{figure*}[t!]
\label{fig:figure1}
\centering
\includegraphics[scale=0.15]{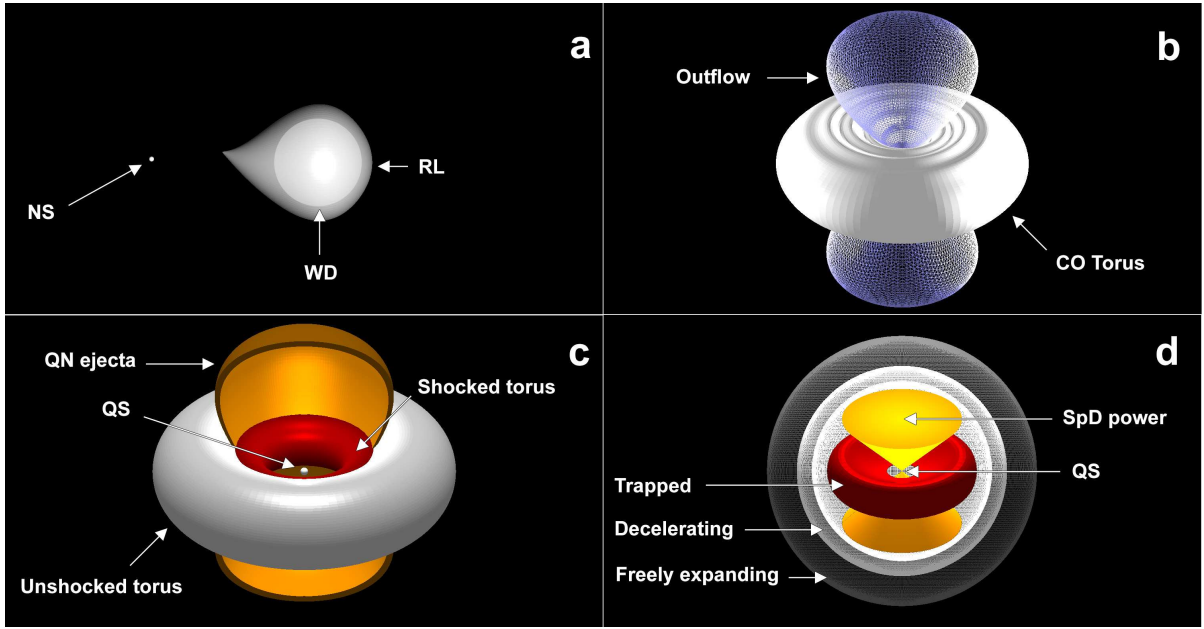}
\caption{Schematic diagram of QN-Ia: 
{\bf Panel a} - A $\sim$0.6 $M_{\odot}$  WD overflows its RL and experiences unstable accretion onto a 1.7$M_{\odot}$ NS.
{\bf Panel b} -  The WD is tidally disrupted and circularizes in a few orbital timescale around the NS forming
a thick CO torus.   {\bf Panel c} -  After accretion of $0.3M_{\odot}$ WD material, the NS experiences an explosive transition to a
 QS (a QN).  The relativistic QN ejecta shocks and  burns the torus. 
 {\bf Panel d} - After the QN shock and ejecta have left the system, some of the burnt torus material remains 
 trapped by the gravitational potential of the QS while some of it is decelerated.  The SpD power from the QS,  
 which contributes to the QN-Ia light-curve, is illustrated as  two  conical lobes.\\}
\end{figure*}

\section{The Quark-Nova Ia model}

  In the QN-Ia, a sub-Chandrasekhar (here $M_{\rm WD, 0} \sim 0.6M_{\odot}$) CO WD overflows its Roche-Lobe (RL) 
 and accretes onto a NS   via a hyper-Eddington accreting  torus (Ouyed \& Staff 2013; Ouyed et al. 2014a).  
  We adopt a critical NS mass of $M_{\rm NS, c.}= 2.0 M_{\odot}$ at which point the    QN is triggered (see Ouyed et al. 2013 for a recent review on the physics of the QN explosion).  
The NS mass at the end of the CE phase is taken to be $M_{\rm NS, 0}= 1.7M_{\odot}$
 which means that for  a  canonical NS birth mass of 1.5$M_{\odot}$ 
   an average of $\sim 0.2 M_{\odot}$ was accreted  during the CE phase (Ouyed et al. 2014a).
        For a typical WD mass of $0.6M_{\odot}$, it means about $0.3M_{\odot}$ needs to be accreted onto the NS 
     to trigger the QN.
 The extremely dense, relativistic, QN ejecta\footnote{A QN ejects $\sim 10^{-3}M_{\odot}$ of neutron-rich material (the outermost layers  of the NS) with a Lorentz factor of about 10 (Ker\"anen et al. 2005).  The resulting QN compact remnant (the QS) has a mass $M_{\rm NS, c.}\sim 2M_{\odot}$.  Such  heavy QSs may exist, so long as the quark 
 superconducting gap and strong coupling corrections are taken into account (e.g., Alford et al. 2007; Buballa et al. 2014).} impacts and shocks the remaining $0.3M_{\odot}$  CO torus, 
 triggering its   thermonuclear burning: the QN-Ia.  
   In addition to the energy from the $^{56}$Ni decay, a QN-Ia is also powered by spin-down (hereafter SpD; to
 be differentiated from SD which stands for Single-Degenerate) from the QS\footnote{A $10^{15}$ G magnetic field can readily be obtained during QS formation due to the response of quarks to the spontaneous magnetization of the gluons (Iwazaki 2005).}.  This results in the QN-Ia obeying a Phillips-like relation where the variation in luminosity is due to SpD power (see Ouyed et al. 2014a for more details).

 Figure 1 (panel a) illustrates the components in the QN-Ia model namely, an accreting NS (which eventually converts
 to a QS),  a WD (the donor) at  an orbital separation  $a=a_{\rm RL}$  (the semi-major axis)   which is
 of the order of a few $10^9$ cm when the WD overflows its RL (using the Eggleton (1983) formula).
    There are two possible outcomes once the WD overflows its RL depending on whether
 mass transfer is stable or unstable (e.g. Verbunt \& Rappaport 1988).  For mass ratio $q=M_{\rm WD, 0}/M_{\rm NS, 0} < \sim 0.3$ accretion proceeds in a stable
  manner  and the WD detonates (following impact by the QN ejecta) while still in orbit
  at $a \sim a_{\rm RL}$ (Ouyed \& Staff 2013; see also Ouyed et al. 2011).  The 
  scenario  which  we consider here with  $q= 0.35$ (panel b in Figure 1), is the unstable mass transfer regime  where the WD is completely disrupted and circularizes around the NS (Fryer et al. 1999).
   As we show below,  we do not expect any WD left when accretion onto the NS takes place.
    In this case, the QN-Ia results from the thermonuclear burning of the CO torus following impact by the QN ejecta
    (panel c in Figure 1).

\subsection{Torus properties}

   The WD disrupts in a few orbital periods (Fryer et al. 1999) on timescales 
  of $\sim 100$ seconds for an orbital separation of  $a_{\rm RL} \sim 3\times 10^9$ cm.
 It circularizes at a radius $R_{\rm circ.} \simeq a_{\rm RL}  (1+q)  \left(0.5-0.227\ln{q}\right)^4\sim 0.4 a_{\rm RL}\sim 10^9$ cm  
(e.g. Shu\&Lubow 1981). The resulting thick torus (with scale-height $h_{\rm torus} \sim 0.5 r_0$;
$r_0$ is the torus co-ordinate) spreads outward and inward on a viscous timescale of a  few $1000$ seconds for  a torus with viscosity parametrized by  $\alpha\sim 0.01$ (Frank et al. 1992). The characteristic  accretion rate,  $\sim M_{\rm WD}/t_{\rm visc.}$, is a few times  $10^{-4}M_{\odot}$ s$^{-1}$. 

The disrupted WD is optically thick and virializes at the circularization radius which yields
a torus temperature, $T_{\rm torus}$,  of the order of a few $10^8$ K (e.g.  Paczy\'nski 1998; assuming the torus
has similar thermal and rotational energy).  The torus average density  ($\sim M_{\rm torus}/R_{\rm circ.}^3$)  is of the order of a few $10^5$ g cm$^{-3}$.   The ignition conditions for nuclear (Carbon) burning  are $T_{\rm i} = 5\times 10^8$ K and $\rho_{\rm i} = 3\times 10^6$ g cm$^{-3}$  (e.g. Ryan \& Norton 2010) so that immediately after the disruption of the WD  the torus is 
unlikely to ignite (mainly because  of the low density). 
  The torus spreads inward and outward on a viscous timescale. If it remains virialized, the temperature increases
 inward as $1/r_0$, so the innermost part of the torus may undergo nuclear burning assuming the density
 increases above $\rho_{\rm i}$.  Without neutrino cooling,  the innermost part of the torus 
  ($r_0 \le \sim 10^8$ cm) could  undergo steady nuclear burning   but it remains to be
  shown whether a detonation is feasible (Fern\'andez \& Metzger 2013).

\section{The low-velocity $^{56}$Ni}

 Once $0.3M_{\odot}$ of material has been accreted (after $\sim1000$ seconds), the NS experiences
 a QN explosion.  The  QN ejecta  and shock  compresses (by a factor of a few 100) and heats (to temperatures exceeding
 $\sim 10^{10}$ K)  the  remaining $\sim 0.3M_{\odot}$ of torus
 material,  which leads to prompt  thermonuclear burning (Ouyed \& Staff 2013).
 The expansion of the burnt  torus is  driven by the energy released at a velocity    $v_{\rm exp., torus} \sim 1.5\times 10^4\ {\rm km\ s}^{-1}$  (Arnett 1982). We assume efficient burning given the high compression and heating of the torus material: 
 the  inner parts are completely burned to $^{56}$Ni, the central part   mostly burned to $^{56}$Ni
  and the outermost low-density layers are burnt to Intermediate-Mass Elements (IMEs).

For an orbit around a point mass, the velocity (from the vis-viva equation; Logsdon 1998) is
$v^2=G M_{\rm QS} (2/r-1/a)$  which means that for  a given orbit  $a = $const., the velocity at a distance $r$ from the QS is
\begin{equation}
v(r)^2 = v(r_0)^2 - v_{\rm esc.}^2  \left(1-\frac{r_0}{r} \right)\ ,
\end{equation}
where $v_{\rm esc.} = \sqrt{2 GM_{\rm QS}/r_0}$ is the escape velocity at radius $r_0$. Here $M_{\rm QS}$ is the QS
mass,  and $G$ the  gravitational constant. 
In the equation above,  $v(r_0)= v_{\rm exp., torus}$ is the initial expansion velocity
 of the burnt torus (CO) material.   
 The solid curve in Figure 2  shows the velocity at infinity ($v_{\rm \infty}$  at $r>> r_0$) for $^{56}$Ni  expanding from an initial radius
     $r_0$.    The asymptotic velocity is reached quickly with values of $< 10^4$ km s$^{-1}$ for  
 $r_0 < \sim 4500$ km while the $^{56}$Ni ejected from higher orbits, $r_0$, 
 retains a velocity close to $v_{\rm exp., torus}$.   The portion of the burnt torus at $r_0 <  r_{\rm 0, esc.} \sim 2500 \ {\rm km}$  remains bound to the QS  since $v_{\rm esc.} > v_{\rm exp., torus}$ (see panel c in Figure 1 for an illustration).       
   
The integrated mass  of the torus can be derived 
 from $M_{\rm torus}(r_0) = \int_{r_{\rm 0, in}}^{r_{\rm 0}}  2\pi \Sigma r_0 dr_0$
with the surface density $\Sigma = \rho_{\rm torus} h_{\rm torus} \propto r_0^{-1/2}$ for $h_{\rm torus} /r_0=0.5$ and a  
   torus density profile  scaling as $\rho_{\rm torus} \propto r_0^{-3/2}$ using hydrostatic equilibrium for a gas with
 an adiabatic index $\gamma=5/3$.   This yields $M_{\rm torus}(r_0)/M_{\rm torus} =  (r_0^{3/2}-r_{\rm 0, in}^{3/2})/((r_{\rm 0, out}^{3/2}-r_{\rm 0, in}^{3/2})$ where $r_{\rm 0, in}$ and $r_{\rm 0, out}$ are the torus inner and outer radii at the time
of the QN explosion; $M_{\rm torus}$ is the total torus mass at the onset of the QN which is $0.3M_{\odot}$
for our fiducial values.   This is shown as the dashed curve in Figure 2. 
About 10\% of the torus material remains trapped by the
QS gravitational potential, $\sim$20\% decelerates to a few $10^3$ km s$^{-1}$ and the remainder 
freely expanding at speeds  close to $v_{\rm exp, torus}$ (i.e. $> 10^4$ km s$^{-1}$). 
The  inner slowly expanding parts are very $^{56}$Ni-rich and the outer fast moving part while mostly burned to $^{56}$Ni, 
 contain some IMEs in  the outermost expanding layers.

 \begin{figure}
\label{fig:vinf}
\centering
\includegraphics[scale=0.65]{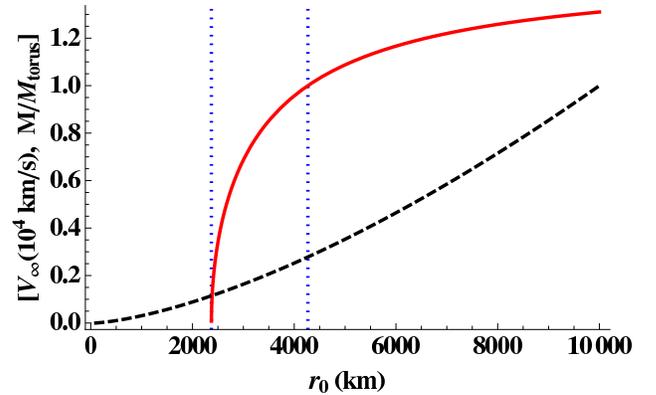}
\caption{The asymptotic velocity ($v_{\rm \infty}$; solid line) of the burnt torus material versus the initial radius $r_0$ for an 
  initial torus expansion velocity  $v_{\rm exp., torus} = 1.5\times 10^4$ km s$^{-1}$. The material within $r_{\rm 0, esc.}\sim 2500$ km (the
  left vertical dotted line)  remains trapped by the QS gravitational potential. The region to the left of the second vertical dotted line (at $r_{\rm 0, esc.}\sim 4500$ km) delineates  the
    region of strongest deceleration with $v_{\rm \infty} < 10^4$ km s$^{-1}$. The dashed line shows the integrated torus mass
    $M_{\rm tours}(r_0)$ un units of the total mass $M_{\rm torus}$.\\}
\end{figure}

\section{The lightcurve (LC)}

 In addition to the energy from the $^{56}$Ni decay, a QN-Ia is also powered by SpD energy of the newly born QS.  
 To compute the QN-Ia LC, we  use the Chatzopoulos et al. (2012) light-curve model which is a generalization of the Arnett (1980 and 1982)
models.  Chatzopoulos et al. (2012) provides formulae for spin-down (their eq. 13) and Nickel-decay (their eq. 9) luminosity in an homologously expanding ejecta.  As explained in the Appendix here, starting with the assumptions of  Chatzopoulos et al. (2012) we make additional assumptions  that
allow us to calculate the QN-Ia lightcurve; we also provide  a link to obtain the code we used.  There are essentially four physical parameters in our model namely:
$M_{\rm eje.}$ (the ejected mass from the torus), $f_{\rm Ni}=M_{\rm Ni} /M_{\rm eje.}$ ($^{56}$Ni fraction of burnt torus), 
         $P_{\rm QS}$  (the QS's initial spin period),    $B_{\rm QS}$ (the QS's  Magnetic field).

Figure 3 shows our best fit to SN 2014J data obtained  by taking a QS with  a period $P_{\rm QS}=18.5$ ms 
and $B_{\rm QS}=1.2\times 10^{15}$ G (which gives SpD timescale $\tau_{\rm SpD}\sim 11.5$   days), and 
 a total ejecta of  $M_{\rm eje.}=0.3 M_{\odot}$. 
The   amount of  $^{56}$Ni  produced (up to $M_{\rm Ni} = M_{\rm eje.}$) for our fiducial values cannot account for the
 luminosity of SN 2014J, in particular at peak,  so that for all cases the LC is also powered by SpD. Using lower  
  Nickel fraction, $f_{\rm Ni}$,  values makes the need for  SpD power even more dominant. 
    Best fits were obtained by  taking  the 
  explosion date relative to the time of peak $m_{\rm B}$ to be at approximately -16.5 days.  Our model is less accurate beyond $\sim$ 30 days past peak because our calculation of the photospheric radius only applies in the optically thick regime (see  Appendix).   The observed early  steep rise of the LC is naturally captured in our model as a consequence  of the SpD power injection.    
     In our model,  QNe-Ia with 5 ms $< P_{\rm QS} <$ 35 ms will display a Phillips relationship (i.e. will be accepted by the LC fitters;  Figure 4 in Ouyed et al. (2014a)).
 Thus the $P_{\rm QS}$ suggested by our fits  means that SN 2014J should obey the Phillips relationship (Phillips 1993).

\begin{figure*}
\label{fig:test1}
\centering
\includegraphics[scale=0.85]{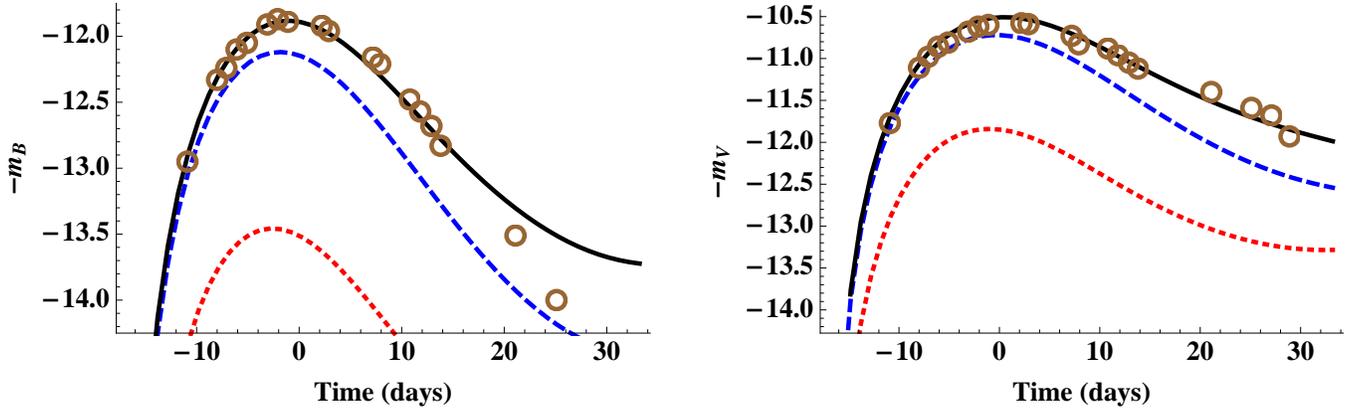}
\caption{Fit to apparent B-band (left panel) and V-band (right panel) SN 2014J data (Marion et al. 2014; time 0
corresponds to peak $m_{\rm B}$). The best fit (solid line) was obtained for ejected mass $M_{\rm eje.}=0.3M_{\odot}$ with  a Nickel fraction of 0.75 and powered
additionally by the SpD power of a  $P_{\rm QS}=18.5$ ms  and $B_{\rm QS}=1.2\times 10^{15}$ G QS (see Table in the Appendix).
 We adopt extinctions of $A_{\rm V}=1.8$ and $A_{\rm B}=3.0$ (Marion et al. 2014). The red-dotted (blue-dashed) line is the pure Nickel (pure SpD) model
with  $M_{\rm eje.}=0.3M_{\odot}$  powered solely by $^{56}$Ni decay (SpD).\\~\\}
\end{figure*}

\section{The optical depth in $\gamma$-rays}

The column density of the spherically outer fast moving  torus material ($\sim$ 70\% of the total ejected mass) is  estimated to be
\begin{equation}
\label{eq:rhocol}
\rho_{\rm col.} \sim \frac{M_{\rm fast}}{4\pi R_{\rm exp., fast} ^2} \sim 13\ {\rm g\ cm}^{-2} \frac{M_{\rm fast, 0.2}}{v_{\rm fast, 10000}^2 t_{20}^2 }\ ,
\end{equation}
with  $R_{\rm exp., fast} = v_{\rm fast} t$;  $v_{\rm fast, 10000}$ is the velocity of the outer expanding torus
 in units of 10000 km s$^{-1}$, $t_{20}$ is time from explosion in units  of 20 days and $M_{\rm fast, 0.2}$
 in units of $0.2M_{\odot}$.

The observed low-velocity $^{56}$Ni decay lines  are from 
  the low-velocity material (the $\sim$ 10\% decelerated part of the torus with a bulk velocity $v_{\rm slow} \sim 5000$ km s$^{-1}$). 
  The average value of $\cos(\theta )$ over a half hemisphere is 0.5 where
$\theta$ is the angle of the velocity of the expanding torus material with respect to the line-of-sight.
  Assuming that the Nickel lines are observed from the  approaching hemisphere, $v_{\rm line} = v_{\rm slow}/2 = 2500$ km s$^{-1}$. 
 Applying equation  \ref{eq:rhocol} to the low-velocity material ($M_{\rm low}\sim 0.1\times 0.3M_{\odot}\sim 0.03M_{\odot}$)
 with $v_{\rm slow}\sim 5000$ km s$^{-1}$ yields  a column density of $\sim 7$ g cm$^{-2}$.
The optical depth of the fast moving part of the ejected torus ($\sim$ 13 g cm$^{-2}$) is  low enough
 that the 158 and 812 keV lines
  from $^{56}$Ni decay  can escape with only a few electron scatterings:  $1/\kappa_{\rm es} \sim 3$ g cm$^{-2}$ where $\kappa_{\rm es}$
is the electron scattering opacity.

 \section{Conclusion and predictions}

The QN-Ia model seems to provide ingredients that can account for the
 kinematics and strength of  $^{56}$Ni decay lines observed in SN 2014J (Diehl et al. 2014a). 
The crucial differences between our model and the standard  SD and DD scenarios 
 are:
 
 (i) The presence of a  gravitational point mass (the QS) which  slows down and traps some of the burnt torus material. 
 In  the SD and DD scenarios a compact remnant may form via  the Accretion-Induced-Channel  channel 
  but in that case no SN-Ia is expected or any significant amount of $^{56}$Ni (Nomoto \& Kondo 1991).

  (ii) The SpD power which provides an additional energy source. 
  One could argue for a similar scenario involving a magnetar (Duncan \& Thompson 1992).  For example, in the scenario
 which has a disrupted WD undergoing detonation (Fern\'andez \& Metzger 2012), the
 resulting ashes can have additional power from the magnetar's SpD.  However, there
 is no apparent mechanism to preserve the magnetic dipole field from the time of  primary star Supernova (i.e.
 Magnetar formation)
 to the WD disruption event, tens of millions of years later. Besides providing  the SpD, the QN provides a means (the QN ejecta) to compress, heat and ignite the torus.

 Our model relies  on the feasibility of the QN explosion.
  Numerical simulations of the burning of a NS to a QS  with consistent 
  treatment of reactions (including neutrinos), diffusion,  and hydrodynamics  find instabilities that  could lead to 
 a detonation  (Niebergal et al. 2010; see also Herzog \& R\"opke 2011).   A ``core-collapse" QN could also result from 
  the collapse of the  strange quark matter core (Ouyed et al. 2013).  
 More sophisticated high-resolution simulations are ultimately required to confirm that the QN is feasible.

\subsection{Predictions}

In the context of SN 2014J, among the 
 predictions that can be tested in the near future  are (see Ouyed et al. 2014b for
 an extended  list) :

\begin{itemize}

\item  Churazov et al. (2014)  measured $^{56}$Ni
mass and velocity which are roughly  in agreement with our model\footnote{Diehl et al. (2014b)  compared the time evolution the measured fluxes of $^{56}$Co  emission to  different models
finding a $^{56}$Ni mass of $\sim 0.49M_{\odot}$.}.  
Churazov et al. (2014)  account for the discrepancy between the $^{56}$Ni measured mass the
 $M_{\rm Ni}\sim 0.77M_{\odot}$ derived from peak luminosity 
 by appealing to optical depth in $\gamma$-rays.   We argue that the discrepancy is due to SpD energy, thus 
   we predict that there will be no substantial increase in $^{56}$Co in future observations.  This prediction provides a crucial   test of  SN 2014J as a QN-Ia.

\item  The fast $^{56}$Ni in Figure 2 should be seen in early ($\sim$ 20 days) spectra as
 broad and  blue-shifted (both by $\sim$ 5\%)  lines.  This might be testable from the early $\gamma$-ray 
 observations of SN 2014J (e.g. Diehl et al. 2014a).

\item The QS' current spin period (6 months after explosion)   is  $\sim$ 40 ms.   The QS is an aligned rotator
(Ouyed et al. 2006) from which X-ray pulsation would be seen  if there is accretion
(e.g. from trapped burnt torus material).

  \item     One could detect  signatures of the Keplerian profile  (double-peaked  lines) 
 from  the ashes of the burnt torus material if some of it remained  trapped ($\sim 0.03M_{\odot}$) by the QS gravitational potential.

\item If the QS turns to a Black Hole early  following the explosion,  this could be observed 
 as a ``glitch" in the LC as the SpD energy is suddenly extinguished (Ouyed et al. 2014b).

\end{itemize}

\begin{acknowledgements}   

This research is supported by  operating grants from the National Science and Engineering Research Council of Canada (NSERC).

\end{acknowledgements}


\begin{appendix}

\section{QN-Ia model LC and associated code}

\subsection{Model assumptions}

We are using the Chatzopoulos et al. (2012) light-curve model which is a generalization of the Arnett (1980 and 1982)
models.  Chatzopoulos et al. (2012) provides formulae for spin-down (their eq. 13) and Nickel-decay (their eq. 9) luminosity in an homologously expanding ejecta.  This means we adopt their assumptions (see Appendix A in Chatzopoulos et al. (2012) and also section II.a in Arnett 1982) which we list here: 

i) Homologously expanding ejecta ($v \propto r$) and a radial density profile given by an arbitrary continuous function $\eta(r)$.

ii) Power source given by eq. A6 in in Chatzopoulos et al. (2012) with radial distribution 
 given by a centrally concentrated function $\zeta(r)$. 

iii) Radiation-pressure dominated with a temperature distribution given by $\psi(r)$ 
with the restriction that   $\zeta(r) \eta(r)/\psi(r)$ is  a constant  (this assumption is required  to obtain the semi-analytical
solutions for the LC).

iv) The output luminosity is the diffusion luminosity as given in eq.(2) in Chatzopoulos et al. (2012).

v) The optical and $\gamma$-ray opacity is taken to be the Thompson opacity of   $0.33$ cm${^2}$ g$^{-1}$ corresponding to a fully ionized solar metallicity material; i.e. the composition is assumed to be fully ionized solar material for the radiative transfer problem.

In addition we add the following assumptions:

vi)  We assume that the radiative transfer solution of Chatzopoulos et al. (2012) for $^{56}$Ni luminosity is not affected by the
SpD luminosity and in turn that the radiative transfer solution for the SpD  luminosity is not affected by the
 $^{56}$Ni luminosity.  This allows us to take the QN-Ia Model total Luminosity  to be additive (i.e. is given by $L_{^{56}Ni} + 
      L_{\rm SpD}$). This is partly justified since in the Chatzopoulos et al. (2012) model  the power input 
      is assumed to not affect the expansion dynamics.   By considering the case of equal SpD and $^{56}$Ni luminosities,
      and using eq. A2 in Chatzopoulos et al. (2012),   we
      estimate an error caused by this assumption of about 20\% or less.

vii)  The ejecta velocity is dominated by the burning of CO to $^{56}$Ni which 
overwhelms either SpD (for $P_{\rm QS} >$ 5 ms) or  $^{56}$Ni decay energy.
This gives an expansion velocity $v_{\rm exp.}\propto f_{\rm Ni}^{1/2}$ (Arnett 1982).

viii)  To get an effective temperature we calculate a photospheric radius $R_{\rm ph.}$ 
for uniform ejecta (at fixed time) and assume a blackbody spectrum. 
     It means   our LC model is applicable only during the optically thick phase.
     The photospheric radius is given by :  $R_{\rm ph.} = 
 R_0 + v_{\rm exp} t - 2 \lambda/3$ with $\lambda$ the photon mean-free-path.
  The second term  includes the first-order effect of inward recession of the
      photosphere; e.g. Arnett 1982).

\subsection{Model  parameters and fitting procedure}

The QN-Ia model has six free parameters:  $M_{\rm eje.}$ (the ejected mass from the torus), $f_{\rm Ni}$ ($^{56}$Ni fraction of the burnt torus), 
         $P_{\rm QS}$  (the QS's initial spin period which affects spin-down luminosity and timescale), 
   $B_{\rm QS}$ (the QS's  initial magnetic field which affects spin-down luminosity and timescale), 
   $\tau_{\rm shift}$  (the model's explosion day relative to time of peak $m_{\rm B}$), and finally 
   $R_0$ (the radius at shock breakout).  The $\tau_{\rm shift}$ is related to the unknown explosion
   date and is not related to the explosion physics. Also, 
    because $R_0$ is 
     negligible for WD explosion models we effectively have four free physical parameters.

     Since the LC is spin-down dominated in our model (because of  the observed low $\gamma$-ray 
optical depth, which means $M_{\rm eje.} \le 0.3M_{\odot}$), $L_{\rm SpD}\propto B_{\rm QS}^2/P_{\rm QS}^4$ is approximately constant when fitting the SN2014J LC;
 best fits are obtained when    $B_{\rm QS, 15} \sim 3.5\times 10^{-3}  P_{\rm QS}^2$, effectively reducing the number of free parameters to three.
 Since the SpD timescale is $\tau_{\rm SpD}\propto P_{\rm QS}^2/B_{\rm QS}^2$ 
 this means that $\tau_{\rm SpD}\propto P_{\rm QS}^{-2}$. In other words, a shorter period
  would give a longer $\tau_{\rm SpD}$ which translates to a slower 
spin-down decay. To compensate one would need a shorter diffusion time ($\tau_{\rm d}\propto M_{\rm eje.}^{1/2}/f_{\rm Ni}^{1/4}$)
  thus  a lower ejecta mass $M_{\rm eje.}$ or a higher $f_{\rm Ni}$; the dependency on $f_{\rm Ni}$ is weaker. Upon experimenting
  with the fits we find that  $f_{\rm Ni}$   decreases to get the ratio of B and V  
magnitudes correct (which we attribute to photospheric  temperature effects).  Once a model is obtained one can shift the explosion date relative to the time of peak $m_{\rm B}$ by $\tau_{\rm shift}$ to overlap with data. A range of acceptable fits are shown in
Table \ref{table:fits}.

 \begin{table}[h!]
\centering
\caption{Examples of acceptable fits using $B_{\rm QS, 15} = 3.5\times 10^{-3} P_{\rm QS}^2$.}
\begin{tabular}{|c|c|c|c|}\hline
$M_{\rm eje.} (M_{\odot})$ & $f_{\rm Ni}$ &  $P_{\rm QS} ({\rm ms})$ &  $\tau_{\rm shift} ({\rm days})$\\\hline
0.30 & 0.75 & 18.5 &16.5 \\
0.25 & 0.65 & 17.5 &16.5 \\
0.20 & 0.55 & 16.5 &16.5 \\\hline
\end{tabular}
\label{table:fits}
\end{table}

\subsection{How to obtain and use the code}

The code and usage instructions are  freely available at: \\
 {\it ftp://quarknova.ucalgary.ca/Codes/QNIa-LC-PACKAGE.zip}.

\end{appendix}

\end{document}